# Adaptive Modified RISE Control for Quadrotors: Enhancing Trajectory Tracking Through Uncertainty Compensation

**KEVIN JOHNSTON[1], (Student Member, IEEE) MUSABBIR AHMED ARRAFI[1], (Student Member, IEEE) KRISHNA B KIDAMBI[1], (Senior Member, IEEE), MADHUR TIWARI[2]**

[1]Dayton Robotics and Controls (DaRC) Lab, Department of Mechanical and Aerospace Engineering, University of Dayton, Dayton, OH 45469 USA ({johnstonk9, arrafim1, kkidambi1}@udayton.edu)

[2]Department of Aerospace, Physics and Space Sciences, Florida Institute of Technology, Melbourne, FL 32901 USA (mtiwari1@fit.edu)

Corresponding author: Krishna B Kidambi (e-mail: kkidambi1@udayton.edu).

**ABSTRACT** This paper presents an adaptive modified Robust Inverse of Signum Error (AM-RISE) control method, which achieves reliable trajectory tracking control for a quadrotor unmanned aerial vehicle. The proposed method systematically accounts for gyroscopic effects, rotor dynamics, parametric uncertainties, and external disturbances, ensuring robust performance across varying trajectory speeds. Through novel mathematical manipulation in the error system development, the quadrotor dynamics are expressed in a control-oriented form, which explicitly incorporates the uncertainty in the gyroscopic term and control actuation term. An adaptive modified RISE law is then designed to stabilize both the position and attitude loops of the quadrotor system. A rigorous Lyapunov-based analysis is utilized to prove asymptotic trajectory tracking, where the region of convergence can be made arbitrarily large through judicious control gain selection. Moreover, the stability analysis formally addresses gyroscopic effects and actuator uncertainty. To illustrate the performance of the control law, comparative numerical simulation results are provided, which demonstrate the improved closed-loop performance achieved under varying levels of parametric uncertainty, disturbance magnitudes and trajectory speeds.

## I. INTRODUCTION

UNMANNED aerial vehicles (UAVs) (or Quadrotors) have the potential to transport people and things to locations not traditionally served by current modes of air transportation. Hundreds of exciting and promising applications in various domains such as air delivery, aerial docking, surveying, agriculture, architecture, security, and video entertainment are being developed [1]–[3]. Developing reliable and safe control algorithms is still a challenging task in a myraid of quadrotor control applications [4]. Linear control approaches such as PID [5] and LQR [6] have been often investigated, but nonlinear control techniques have been widely shown to achieve an increased level of reliability in quadrotor flight under various uncertain and adversarial operating conditions. Common nonlinear control design approaches include feedback linearization [7], backstepping [8], [9], sliding mode [10], [11], adaptive control [12], and robust control [13], [14]. Learning-based controllers have gained significant attention, however they often struggle to generalize effectively beyond their training data and the generation of diverse and sufficient training data remains an open challenge [15].

Advantages to feedback linearization include mathematical simplicity and ease of implementation; however, they often incur reduced robustness to disturbances and sensitivity to model uncertainty. The authors in [16] present an adaptive back-stepping controller to track time-varying trajectories with parameter uncertainties and under-actuation, but their approach ignores the aerodynamic/propeller effects. Nonlinear adaptive control methods have also been presented to compensate for disturbances, but these results can often be limited to local convergence results [17], [18]. Sliding mode control approaches are widely known for their disturbance-rejection capabilities, but their implementation can be challenging due to the use of high-bandwidth switching [19]. The authors in [20] develop an agile control subjected to aerodynamic drag, center of gravity shift, and motor dynamics. It is shown that accounting for uncertainties in the actuator dynamics is essential to achieve the desired trajectory-tracking performance. In [21], an Unscented Kalman Filter (UKF) is designed to estimate a quadrotor's inertial and aerodynamic parameters, linear velocity, angular velocity, and tilt using only IMU and motor speed measurements, thereby eliminating the reliance on external sensors such as motion capture systems during trajectory tracking.

Despite these advancements, ensuring robustness for quadrotor control under plethora of unmodeled disturbances, especially when explicitly incorporating complex physical effects and actuator dynamics remains an active area of research [22]. Motivated by this, robust integral of the signum of the error (RISE)-based control can be used to achieve reliable disturbance rejection using continuous control [23]. RISE has been widely shown to achieve superior rejection of norm-bounded disturbances in various recent results [24], [25]. In addition, our recent result in [26] presents a modified RISE (MRISE)-based quadrotor control design, which is proven to reject external disturbances and compensate for varying levels of parametric uncertainties in the UAV dynamics. While the MRISE result in [26] achieved significant performance improvement in trajectory tracking compared to traditional RISE, the work does not formally incorporate the effects of actuator dynamics, gyroscopic, and aerodynamic/propeller effects into the quadrotor dynamic model.

Motivated by this and future demands of building robust and agile quadrotors, we propose an adaptive modified robust inverse of signum error (AM-RISE) based control method for a quadrotor trajectory tracking subjected to parametric uncertainty, unmodeled gyroscopic, aerodynamic effects and uncertainties in the actuator dynamics in addition to external disturbances. The primary contributions of this result can be summarized as follows:

1) An augmented AM-RISE control design for both the inner and outer loop of the quadrotor tracking control system, which formally incorporates feedback compensators for actuator uncertainty, gyroscopic, aero effects, model uncertainties, and external disturbances. The control algorithm accounts for varying magnitudes of parameteric uncertainty ($-25\%$ to $30\%$) with little or no gain tuning.
2) The proposed control algorithm demonstrates robust performance by successfully tracking trajectories with varying speeds (1 to 4 $\frac{m}{s}$) and under varying disturbance magnitudes ($-2$ to 2 $\frac{m}{s}$).
3) A rigorous closed-loop stability analysis using a Lyapunov-based approach provides a detailed derivation of semi-global asymptotic tracking and the specific control gain conditions under which convergence can be guaranteed.
4) A detailed comparative numerical simulation study was conducted to demonstrate the performance of the proposed control design.

This work extends our preliminary study presented in [27], where an adaptive modified RISE-based controller was developed for quadrotor trajectory tracking under actuator uncertainty. The present version significantly advances that work through comprehensive simulation updates (covering variation of uncertainty in system model, actuator dynamics (see Table.2), and varying magnitudes of trajectory speeds, and external disturbances), expanded theoretical analysis, and the inclusion of two additional co-authors who contributed to the extended controller design, implementation, and validation.

The rest of the paper is organized as follows: Section II describes the nonlinear dynamic model of the quadrotor with gyroscopic effects, especially Sections II-B describes a detailed model incorporating actuator dynamics. The open-loop and closed-loop error systems are described in Section III-A and III-B, respectively. In Section IV, the closed-loop stability analysis is presented. Finally, Section V presents the numerical simulations to demonstrate the effectiveness of the proposed method, followed by conclusions in Section VI.

## II. MATHEMATICAL MODEL

### A. QUAD DYNAMICS

The mathematical model of quadrotor dynamics is well studied in the literature [28], [29]. The explicit form of the dynamics by using the Newton-Euler method with position and orientation can be written as [21]

$$m\dot{V} = F_g + F_T + F_d + d_1(t), \tag{1a}$$

$$I\dot{\Omega} = -\Omega^{\times} I\Omega + \tau - \tau_a + \tau_g + d_2(t) \tag{1b}$$

where $m \in \mathbb{R}^+$ is the mass of the quadrotor, $V = [V_x, V_y, V_z]^T$ and $\Omega = [p, q, r]^T$ denote the translational and angular velocity defined in the body frame. $I = diag(I_x, I_y, I_z) \in \mathbb{R}^{3\times 3}$ is a positive definite diagonal moment of inertia matrix, and $d_1(t), d_2(t)$ are the external disturbances in the translation and rotational dynamics respectively. The remaining forces in the translational dynamics are given as $F_g = [0, 0, -mg]^T$, where $g = 9.81\ \frac{m}{s^2}$ is the gravitational acceleration. $F_d$ is the drag due to the translational motion given by

$$F_d = -diag(K_{d_x}, K_{d_y}, K_{d_z})V \tag{2}$$

where $K_{d_x}, K_{d_y}, K_{d_z}$ are translational drag coefficients.

The control input $F_T \in \mathbb{R}$ is the total thrust generated by all four motors and is given as

$$F_T = R_1 \sum_{i=1}^{4} F_i \tag{3}$$

where $F_i = K_T\omega_i^2$, and $K_T$ is the thrust coefficient and $\omega_i$ is the speed of $i^{th}$ motor. The term $R_1$ is given as

$$R_1 = \begin{bmatrix} \cos(\psi)\sin(\theta)\cos(\phi) + \sin(\psi)\sin(\phi) \\ \sin(\phi)\sin(\theta)\sin(\psi) - \cos(\psi)\sin(\phi) \\ \cos(\theta)\cos(\phi) \end{bmatrix}. \tag{4}$$

The gyroscopic torque $\tau_g$ is given as

$$\tau_g = \sum_{i=1}^{4} \Omega^{\times} I_r \left[0, 0, (-1)^{i-1}\omega_i\right]^T \tag{5}$$

where $I_r$ is the rotor inertia.

The aerodynamic frictional torque $\tau_a$ is given as [30]

$$\tau_a = -diag(K_{a_x}, K_{a_y}, K_{a_z})\|\Omega^2\| \tag{6}$$

where $K_{a_x}, K_{a_y}, K_{a_z}$ are the coefficients of aerodynamic friction.

The torque $\tau$ applied on the quadrotor is given by

$$\tau = \begin{bmatrix}\tau_\phi\\ \tau_\theta\\ \tau_\psi\end{bmatrix} = \begin{bmatrix}0 & -lK_T & 0 & lK_T\\ -lK_T & 0 & lK_T & 0\\ K_D & -K_D & K_D & -K_D\end{bmatrix}\begin{bmatrix}\omega_1^2\\ \omega_2^2\\ \omega_3^2\\ \omega_4^2\end{bmatrix} \tag{7}$$

where $\tau_\phi, \tau_\theta, \tau_\psi$ denotes the torque along $x, y, z$ axis respectively, $l$ is the distance between each rotor to the quadrotor center of mass, $K_T$ is the thrust factor and $K_D$ is the drag coefficient.

### B. ACTUATOR DYNAMICS

The control input commands to the quadrotor are applied through a DC motor and when tracking aggressive trajectories, actuator dynamics cannot be ignored for the sake of precise tracking performance. The dynamics of the $i^{th}$ motor speed is given as [20], [21]

$$\dot{\omega}_i = \frac{k_\tau}{I_r}(\omega_{ci} - \omega_i) \tag{8}$$

where $k_\tau$ is the motor coefficient and $\omega_{ci}$ is the commanded speed of the $i^{th}$ motor, and $\omega_i$ is introduced in (3).

### C. UNCERTAIN QUADROTOR MODEL

The dynamic equations in (1) are recast by incorporating parametric and actuator uncertainties and are simplified as

$$\dot{P} = V \tag{9a}$$
$$\dot{V} = f_1(V) + F_V + \Delta_V F_V + d_V(t) \tag{9b}$$
$$\dot{R}_1 = R_1[\Omega]^\times \tag{9c}$$
$$\dot{\Omega} = f_2(\Omega) + \tau_\Omega + \Delta_\Omega \tau_\Omega + d_\Omega(t) \tag{9d}$$

where $P = [x,\ y,\ z]^T$, $V = [V_x,\ V_y,\ V_z]^T$, $\Theta = [\phi,\ \theta,\ \psi]^T$ and $\Omega = [p, q, r]^T$ are the position, velocity, angular position, and angular velocity in 3-D. The notation $[\cdot]^\times$ represents the skew-symmetric matrix operator for a given vector, defined as:

$$[\Omega]^\times = \begin{bmatrix}0 & -r & q\\ r & 0 & -p\\ -q & p & 0\end{bmatrix}. \tag{10}$$

The following definitions have been made

$$F_V = \frac{F_T}{m} \tag{11a}$$
$$\Delta_V F_V = \frac{R_1}{m}\sum_{i=1}^{4}\Delta F_i \tag{11b}$$
$$f_1(V) = \frac{\bar{F}_g}{m} + \frac{\bar{F}_d}{m} \tag{11c}$$
$$d_V(t) = -\frac{\Delta m\dot{V}}{m} + \frac{\Delta F_g}{m} + \frac{\Delta F_d}{m} + \frac{d_1(t)}{m} \tag{11d}$$
$$d_1 = dis_{amp} * [\sin(t); cos(t); \sin(t) * \cos(t)]^T \tag{11e}$$

$$\tau_\Omega = R_1 \cdot [u_1^2\ u_2^2\ u_3^2\ u_4^2]^T \tag{12a}$$
$$\Delta_\Omega \tau_\Omega = \Delta R_1 \cdot [u_1^2\ u_2^2\ u_3^2\ u_4^2]^T \tag{12b}$$
$$f_2(\Omega) = I^{-1}\Big[-\overline{\Omega^\times I\Omega} - \bar{\tau}_a + \bar{\tau}_g\Big] \tag{12c}$$
$$d_\Omega(t) = I^{-1}\Big[-\overline{\Omega^\times \Delta I\Omega} - \Delta I\dot{\Omega} - \Delta\tau_a + \Delta\tau_g + d_2(t)\Big] \tag{12d}$$
$$d_2 = dis_{amp}\begin{bmatrix}\sin(0.5t)\\ \cos(0.5t)\\ \sin(0.5t)\cdot\cos(0.5t)\end{bmatrix}. \tag{12e}$$

## III. CONTROL MODEL

The control objective is to design the control signal to regulate the quadrotor states to a given desired time-varying reference trajectory that is sufficiently smooth despite the presence of uncertainties in the dynamic and actuator model respectively. Thus, the control objective can be mathematically stated as

$$\|\xi(t)\| \to 0 \tag{13}$$

where $\xi$ in (13) is the difference between the current state and desired state of the quadrotor and $\|(.)\|$ in (13) denotes standard Euclidean norm.

*Assumption 1:* The desired trajectory profile $P_d(t)$ and its first three-time derivatives are bounded in the sense that $P_d(t), \dot{P}_d(t), \ddot{P}_d(t), \dddot{P}_d(t) \in \mathcal{L}_\infty \forall\, t \geq 0$.

*Assumption 2:* The disturbances in the quadrotor dynamics in (9b) and (9d) are smooth enough such that

$$\|\dot{d}_V\| \leq \epsilon_1, \qquad \|\dot{d}_\Omega\| \leq \epsilon_2 \tag{14}$$

where $\epsilon_1, \epsilon_2$ are some unknown positive constants.

### A. OPEN-LOOP ERROR SYSTEM

#### 1) Position Loop

To assess the controller performance and to enable the subsequent stability analysis, the tracking error $\xi_1(t) \in \mathbb{R}^3$ and auxiliary tracking errors $\xi_2(t), \eta_1(t) \in \mathbb{R}^3$ for the position loop are defined as

$$\xi_1 = P - P_d, \qquad \xi_2 = \dot{\xi}_1 + k_1\xi_1, \tag{15}$$
$$\eta_1 = \dot{\xi}_2 + k_2\xi_2 \tag{16}$$

where $P_d = [x_d,\ y_d,\ z_d]^T \in \mathbb{R}^3$, $V_d = [V_{d_x},\ V_{d_y},\ V_{d_z}]^T \in \mathbb{R}^3$ are the desired position and velocity; and $k_1, k_2 \in \mathbb{R}$ denote positive control gains.

Taking the derivative of (16) and using (15) and (9b), the open loop error dynamics can be expressed as

$$\dot{\eta}_1 = \dot{f}_1(V) + \dot{F}_V(t) + \dot{\Delta}_V F_V(t) + \dot{d}_V(t) - \ddot{V}_d + k_1(\dot{\xi}_2 - k_1\dot{\xi}_1) + k_2\dot{\xi}_2. \tag{17}$$

The error dynamics in (17) can be expressed as

$$\dot{\eta}_1 = \tilde{\Gamma}_1 + \Gamma_{1d} + \dot{F}_V(t) + \dot{\Delta}_V F_V(t) - \xi_2 \tag{18}$$

where the unknown auxiliary functions, $\tilde{\Gamma}_1(t), \Gamma_{1d}(t) \in \mathbb{R}^3$ are defined as

$$\tilde{\Gamma}_1(t) = \dot{f}_1(V) + k_1(\dot{\xi}_2 - k_1\dot{\xi}_1) + k_2\dot{\xi}_2 + \xi_2, \tag{19}$$
$$\Gamma_{1d} = \dot{d}_V(t) - \ddot{V}_d. \tag{20}$$

The motivation for the separation of terms in (19) and (20) is based on the fact that the following inequalities can be developed

$$\|\tilde{\Gamma}_1\| \leq \rho_1 \left(\|\mu_1\|\right) \|\mu_1\|, \tag{21}$$

$$\|\Gamma_{1d}\| \leq \kappa_{\Gamma_{1d}}, \quad \|\dot{\Gamma}_{1d}\| \leq \kappa_{\dot{\Gamma}_{1d}} \tag{22}$$

where $\kappa_{\Gamma_{1d}}, \kappa_{\dot{\Gamma}_{1d}} \in \mathbb{R}^+$ are known bounding constants; $\rho(\cdot)$ is a positive, globally invertible, non-decreasing function; and $\mu_1(t) \in \mathbb{R}^9$ is defined as

$$\mu_1(t) \triangleq \left[\begin{array}{ccc} \xi_1^T(t) & \xi_2^T(t) & \eta_1^T(t) \end{array}\right]^T. \tag{23}$$

### 2) Attitude Loop

The tracking error in the attitude loop $\xi_3(t) \in \mathbb{R}^3$ and auxiliary tracking errors $\xi_4(t), \eta_2(t) \in \mathbb{R}^3$ are defined as

$$\xi_3 = \omega - \omega_d, \qquad \xi_4 = \dot{\xi}_3 + k_3\xi_3, \tag{24}$$

$$\eta_2 = \dot{\xi}_4 + k_4\xi_4 \tag{25}$$

where $\omega_d = [\phi_d,\ \theta_d,\ \psi_d]^T \in \mathbb{R}^3, \Omega_d = [p_d,\ q_d,\ r_d]^T \in \mathbb{R}^3$ are the desired angular position and angular velocity, $k_3, k_4 \in \mathbb{R}$ denote positive control gains.

Since the quadrotor is an underactuated system with four control inputs and six outputs, using the differential flatness property [31], the control input can be obtained from four flat outputs $[x_d,\ y_d,\ z_d,\ \psi_d]^T$ (user-defined). In addition, the desired trajectories of roll ($\phi_d$) and pitch ($\theta_d$) depend on these flat outputs (cf. Eq. (19), (20) and (21) in [26]).

By taking the derivative of (25) and using (24), (9d), the open loop error dynamics in the attitude loop can be expressed as

$$\begin{aligned}\dot{\eta}_2 =& \dot{f}_2(\Omega) + \dot{\tau}_\Omega(t) + \dot{\Delta}_\Omega\tau_\Omega(t) + \dot{d}_\Omega(t) - \ddot{\Omega}_d \\ &+ k_3(\dot{\xi}_4 - k_3\dot{\xi}_3) + k_4\dot{\xi}_4.\end{aligned} \tag{26}$$

The error dynamics in (26) can be expressed as

$$\dot{\eta}_2 = \tilde{\Gamma}_2 + \Gamma_{2d} + \dot{\tau}_\Omega(t) + \dot{\Delta}_\Omega\tau_\Omega(t) - \xi_4 \tag{27}$$

where the unknown auxiliary functions $\tilde{\Gamma}_2(t), \Gamma_{2d}(t) \in \mathbb{R}^3$ are defined as

$$\tilde{\Gamma}_2(t) = \dot{f}_2(\Omega) + k_3(\dot{\xi}_4 - k_3\dot{\xi}_3) + k_4\dot{\xi}_4 + \xi_4, \tag{28}$$

$$\Gamma_{2d} = \dot{d}_\Omega(t) - \ddot{\Omega}_d. \tag{29}$$

*Assumption 3:* Approximate model knowledge is available such that $\dot{\Delta}_V(t)$ and $\dot{\Delta}_\Omega(t)$ satisfies

$$\left\|\dot{\Delta}_V(t)\right\|_{i\infty} < \delta_1 < 1, \qquad \left\|\dot{\Delta}_\Omega(t)\right\|_{i\infty} < \delta_2 < 1 \tag{30}$$

where $\delta_1, \delta_2 \in \mathbb{R}^+$ is a known bounding constant, and $\|(\cdot)\|_{i\infty}$ denotes the induced infinity norm.

The motivation for the separation of terms in (28) and (29) are based on the fact that the following inequalities can be developed

$$\|\tilde{\Gamma}_2\| \leq \rho_2 \left(\|\mu_2\|\right) \|\mu_2\|, \tag{31}$$

$$\|\Gamma_{2d}\| \leq \kappa_{\Gamma_{2d}}, \quad \|\dot{\Gamma}_{2d}\| \leq \kappa_{\dot{\Gamma}_{2d}} \tag{32}$$

where $\kappa_{\Gamma_{2d}}, \kappa_{\dot{\Gamma}_{2d}} \in \mathbb{R}^+$ are known bounding constants; $\rho(\cdot)$ is a positive, globally invertible, non-decreasing function; and $\mu_2(t) \in \mathbb{R}^9$ is defined as

$$\mu_2(t) \triangleq \left[\begin{array}{ccc} \xi_3^T(t) & \xi_4^T(t) & \eta_2^T(t) \end{array}\right]^T. \tag{33}$$

## B. CLOSED-LOOP ERROR SYSTEM

### 1) Position Loop

Based on the open-loop error system dynamics in (18), the control term $F_V(t)$ is designed as

$$\dot{F}_V(t) = -\alpha_F\|F_V(t)\|sgn(\eta_1) - (\alpha_1+1)\eta_1 - \hat{\lambda}_1 sgn(\eta_1) \tag{34}$$

where $\alpha_F, \alpha_1 \in \mathbb{R}$ are positive control gains. $\hat{\lambda}_1 \in \mathbb{R}^3$ is an adaptive nonlinear feedback gain, which is defined as

$$\dot{\hat{\lambda}}_{1i} = \beta_1 sgn(\eta_1)\eta_1 \tag{35}$$

where $i = 1, 2, 3$, $\beta_1$ is a positive adaptation gain.

After substituting (34) into (18), the closed-loop error dynamics are obtained as

$$\begin{aligned}\dot{\eta}_1 =& \tilde{\Gamma}_1 + \Gamma_{1d} - \alpha_F\|F_V(t)\|sgn(\eta_1) - \xi_2 \\ &- (\alpha_1+1)\eta_1 - \hat{\lambda}_1 sgn(\eta_1) + \dot{\Delta}_V F_V(t).\end{aligned} \tag{36}$$

### 2) Attitude Loop

Based on the open-loop error system dynamics in (27), the control term $\tau_\Omega(t)$ is designed as

$$\dot{\tau}_\Omega(t) = -\alpha_\Omega\|\tau_\Omega(t)\|sgn(\eta_2) - (\alpha_2+1)\eta_2 - \hat{\lambda}_2 sgn(\eta_2) \tag{37}$$

where $\alpha_\Omega, \alpha_2 \in \mathbb{R}$ are positive, control gains. $\hat{\lambda}_2 \in \mathbb{R}^3$ is a adaptive nonlinear feedback gain, which is defined as

$$\dot{\hat{\lambda}}_{2i} = \beta_2 sgn(\eta_2)\eta_2 \tag{38}$$

where $i = 1, 2, 3$; $\beta_2$ is a positive adaptation gain.

After substituting (37) into (27), the closed-loop error dynamics are obtained as

$$\begin{aligned}\dot{\eta}_2 =& \tilde{\Gamma}_2 + \Gamma_{2d} - \alpha_\Omega\|\tau_\Omega(t)\|sgn(\eta_2) - \xi_4 \\ &- (\alpha_2+1)\eta_2 - \hat{\lambda}_2 sgn(\eta_2) + \dot{\Delta}_\Omega\tau_\Omega(t).\end{aligned} \tag{39}$$

# IV. STABILITY ANALYSIS

Before providing the stability analysis of the closed-loop quadrotor system, the following conditions and definition will be utilized in the proof of Theorem 1.

*Condition 1 ($\lambda_{ji}$-Gain):* To facilitate the following stability proof, the adaptive gain $\lambda_{ji}$ for $j = 1, 2$ and $i = 1, 2, 3$ must satisfy the following condition

$$\lambda_{1i} = \Gamma_{1d} + \frac{1}{k_2}\dot{\Gamma}_{1d}, \qquad \lambda_{2i} = \Gamma_{2d} + \frac{1}{k_4}\dot{\Gamma}_{2d}. \tag{40}$$

*Definition 1:* To facilitate the Lyapunov-based proof of Theorem 1, two functions $Q_1(t), Q_2(t) \in \mathbb{R}$ are defined as

$$Q_1(t) \triangleq \lambda_1|\xi_2(0)| - \xi_2^T(0)\Gamma_{1d}(0) - \int_0^t R_1(v)dv, \tag{41}$$

$$Q_2(t) \triangleq \lambda_2|\xi_4(0)| - \xi_4^T(0)\Gamma_{2d}(0) - \int_0^t R_2(v)dv \tag{42}$$

where the auxiliary functions $R_1(t), R_2(t) \in \mathbb{R}$ are defined as

$$R_1(t) = \eta_1^T(t)\left[\Gamma_{1d}(t) - \lambda_1 sgn(\eta_1)\right], \quad (43)$$
$$R_2(t) = \eta_2^T(t)\left[\Gamma_{2d}(t) - \lambda_2 sgn(\eta_2)\right]. \quad (44)$$

*Condition 2 (Control Gain):* The control gains defined in (34) and (37) are selected according to the conditions

$$\alpha_1 > \frac{\rho^2(\|\mu_1\|)}{4min\{(k_1 - \frac{1}{2}), (k_2 - \frac{1}{2}), 1\}}, \quad (45)$$

$$\alpha_2 > \frac{\rho^2(\|\mu_2\|)}{4min\{(k_3 - \frac{1}{2}), (k_4 - \frac{1}{2}), 1\}}, \quad (46)$$

$$\alpha_F \geq \delta_1, \qquad \alpha_\tau \geq \delta_2. \quad (47)$$

*Theorem 1:* For the quadrotor dynamics described in (9a)-(9d) and for a given sufficiently smooth desired trajectory $[x_d, y_d, z_d, \psi_d]^T$, the robust nonlinear control law given in (34) and (37) along with the adaptive parameters updated based on (35) and (38) ensure the tracking error is asymptotically regulated and all the signals remain bounded throughout the closed-loop operation in the sense that

$$\|\xi(t)\| \rightarrow 0 \qquad \text{for} \qquad t \geq t_n < \infty \quad (48)$$

where $t_n \in \mathcal{L}_\infty$.
*Proof 1:* Let $\mathcal{K} \subset \mathbb{R}^{6n+6+2}$ be a domain containing $w(t) = 0$, where $w(t) \in \mathbb{R}^{6n+6+2}$ is defined as

$$w(t) \triangleq \left[\begin{array}{ccccc} \mu^T(t) & \tilde{\lambda}_1^T & \tilde{\lambda}_2^T & \sqrt{Q_1(t)} & \sqrt{Q_2(t)} \end{array}\right]^T \quad (49)$$

where $\tilde{\lambda}_i = \hat{\lambda}_i - \lambda_i, (i = 1, 2)$ in (49) represent the difference between the estimated parameter and its true value defined in Condition 1 and $\mu(t) \in \mathbb{R}^{6n}$ is defined as

$$\mu(t) \triangleq \left[\begin{array}{cc} \mu_1^T(t) & \mu_2^T(t) \end{array}\right]^T \quad (50)$$

$Q_1(t)$ and $Q_2(t)$ are defined in Definition 1. Let $V(w,t)$ : $\mathcal{K} \times [0,\infty) \rightarrow \mathbb{R}$ be a radially unbounded, positive definite function defined as

$$\begin{aligned} V(w,t) = &\frac{1}{2}\xi_1^T\xi_1 + \frac{1}{2}\xi_2^T\xi_2 + \frac{1}{2}\xi_3^T\xi_3 + \frac{1}{2}\xi_4^T\xi_4 \\ &+ \frac{1}{2}\eta_1^T\eta_1 + \frac{1}{2}\eta_2^T\eta_2 + Q_1 + Q_2 \\ &+ \frac{1}{2\beta_1}\tilde{\lambda}_1^T\tilde{\lambda}_1 + \frac{1}{2\beta_2}\tilde{\lambda}_2^T\tilde{\lambda}_2. \end{aligned} \quad (51)$$

The expression in (51) satisfies the inequalities

$$U_1(w) \leq V(w,t) \leq U_2(w) \quad (52)$$

provided the control gains in **Condition 1** are satisfied.

In (52), the continuous positive definite functions $U_1(w)$, $U_2(w) \in \mathbb{R}$ are defined as

$$U_1(w) \triangleq \tfrac{1}{2}\|w\|^2, \qquad U_2(w) \triangleq \|w\|^2. \quad (53)$$

After taking the time derivative of (51) and using (15), (24), (36), (39), $\dot{V}$ can be expressed as

$$\begin{aligned} \dot{V} = &\xi_1^T[\xi_2 - k_1\xi_1] + \xi_2^T[\eta_1 - k_2\xi_2] + \xi_3^T[\xi_4 - k_3\xi_3] \\ &+ \xi_4^T[\eta_2 - k_4\xi_4] + \eta_1^T[\tilde{\Gamma}_1 + \Gamma_{1d} - \alpha_F\|F_V\|sgn(\eta_1) \\ &- (\alpha_1+1)\eta_1 - \hat{\lambda}_1 sgn(\eta_1) + \dot{\Delta}_V F_V(t) - \xi_2] \\ &+ \eta_2^T[\tilde{\Gamma}_2 + \Gamma_{2d} - \alpha_\Omega\|\tau_\Omega\|sgn(\eta_2) - (\alpha_2+1)\eta_2 \\ &- \hat{\lambda}_2 sgn(\eta_2) + \dot{\Delta}_\Omega\tau_\Omega(t) - \xi_4] + \dot{Q}_1 + \dot{Q}_2 \\ &+ \frac{1}{\beta_1}\tilde{\lambda}_1^T\dot{\hat{\lambda}}_1 + \frac{1}{\beta_2}\tilde{\lambda}_2^T\dot{\hat{\lambda}}_2. \end{aligned} \quad (54)$$

By using the bounding inequalities in (21), (31) and the defition of $Q_1, Q_2$ from (41), (42), and the definition of $\dot{\hat{\lambda}}_1, \dot{\hat{\lambda}}_2$ from (35), (38), the expression in (54) can be upper bounded as

$$\begin{aligned} \dot{V} \leq\ &\xi_1^T\xi_2 + \xi_3^T\xi_4 - k_1\|\xi_1\|^2 - k_2\|\xi_2\|^2 - k_3\|\xi_3\|^2 \\ -\ &k_4\|\xi_4\|^2 + \|\eta_1\|\rho(\|\mu_1\|)\|\mu_1\| + \|\eta_1\|\|\Gamma_{1d}\| \\ -\ &\|\eta_1\|(\alpha_F)\|F_V\| - (\alpha_1+1)\|\eta_1\|^2 - \|\eta_1\|\hat{\lambda}_1 sgn(\eta_1) \\ +\ &\delta_1\|\eta_1\|\|F_V\| + \|\eta_2\|\rho(\|\mu_2\|)\|\mu_2\| + \|\eta_2\|\|\Gamma_{2d}\| \\ -\ &\|\eta_2\|(\alpha_\tau)\|\tau_\Omega\| - (\alpha_2+1)\|\eta_2\|^2 - \|\eta_2\|\|\hat{\lambda}_2 sgn(\eta_2) \\ +\ &\delta_2\|\eta_2\|\|\tau_\Omega\| - \|\eta_1\|\|\Gamma_{1d}\| + \|\eta_1\|\lambda_1 sgn(\eta_1) \\ -\ &\|\eta_2\|\|\Gamma_{2d}\| + \|\eta_2\|\lambda_2 sgn(\eta_2) + \tilde{\lambda}_1^T sgn(\eta_1)\eta_1 \\ +\ &\tilde{\lambda}_2^T sgn(\eta_2)\eta_2. \end{aligned} \quad (55)$$

Young's inequality can be used to upper bound $\xi_1^T\xi_2$ and $\xi_3^T\xi_4$ as

$$\xi_1^T\xi_2 \leq \frac{1}{2}\|\xi_1\|^2 + \frac{1}{2}\|\xi_2\|^2, \quad (56)$$
$$\xi_3^T\xi_4 \leq \frac{1}{2}\|\xi_3\|^2 + \frac{1}{2}\|\xi_4\|^2. \quad (57)$$

By using (56), and the gains $\alpha_F, \alpha_\tau, \alpha_1$, and $\alpha_2$ satisfy the gain condition in Condition 2, the upper bound in (55) can be further simplified as

$$\begin{aligned} \dot{V} \leq\ &-(k_1 - \frac{1}{2})\|\xi_1\|^2 - (k_2 - \frac{1}{2})\|\xi_2\|^2 - \|\eta_1\|^2 - \alpha_1\|\eta_1\|^2 \\ +\ &\rho(\|\mu_1\|)\|\mu_1\|\|\eta_1\| - [\alpha_F - \delta_1]\|\eta_1\|\|F_V\| \\ -\ &(k_3 - \frac{1}{2})\|\xi_3\|^2 - (k_4 - \frac{1}{2})\|\xi_4\|^2 - \|\eta_2\|^2 - \alpha_2\|\eta_2\|^2 \\ +\ &\rho(\|\mu_2\|)\|\mu_2\|\|\eta_2\| - [\alpha_\tau - \delta_2]\|\eta_2\|\|\tau_\Omega\| \end{aligned} \quad (58)$$

$$\begin{aligned} \dot{V} \leq\ &-(k_1 - \frac{1}{2})\|\xi_1\|^2 - (k_2 - \frac{1}{2})\|\xi_2\|^2 - \|\eta_1\|^2 \quad (59) \\ -\ &\alpha_1\left[\|\eta_1\| - \frac{\rho(\|\mu_1\|)}{2\alpha_1}\|\mu_1\|\right]^2 + \frac{\rho^2(\|\mu_1\|)}{4\alpha_1}\|\mu_1\|^2 \\ -\ &(k_3 - \frac{1}{2})\|\xi_3\|^2 - (k_4 - \frac{1}{2})\|\xi_4\|^2 - \|\eta_2\|^2 \\ -\ &\alpha_2\left[\|\eta_2\| - \frac{\rho(\|\mu_2\|)}{2\alpha_2}\|\mu_2\|\right]^2 + \frac{\rho^2(\|\mu_2\|)}{4\alpha_2}\|\mu_2\|^2 \end{aligned}$$

$$\dot{V} \leq -\left[c_1 - \frac{\rho^2(\|\mu\|)}{4c_2}\right]\|\mu\|^2 \quad (60)$$

where $c_1 \triangleq \min\left[min\{(k_1 - \frac{1}{2}), (k_2 - \frac{1}{2}), 1\}, min\{(k_3 - \frac{1}{2}), (k_4 - \frac{1}{2}), 1\}\right]$ and $c_2 \triangleq min\{\alpha_1, \alpha_2\}$.

The following expression can be obtained from (60)

$$\dot{V} \leq -U(w) \tag{61}$$

where $U(w) = c\|\mu\|^2$, for some positive constant $c \in \mathbb{R}$ is a continuous positive semi-definite function that is defined on the domain

$$\mathcal{K} \triangleq \left\{w(t) \in \mathbb{R}^{6n+2} | \, \|w\| \leq \rho^{-1}\left(2\sqrt{c_1 c_2}\right)\right\}. \tag{62}$$

The expressions in (52) and (60), along with (45), (46) can be used to prove that $V(w,t) \in \mathcal{L}_\infty$ in $\mathcal{K}$; hence, $\xi_1(t), \xi_2(t), \xi_3(t), \xi_4(t), \eta_1(t), \eta_2(t) \in \mathcal{L}_\infty$ in $\mathcal{K}$. Given that $\xi_1(t), \xi_2(t), \xi_3(t), \xi_4(t), \eta_1(t), \eta_2(t) \in \mathcal{L}_\infty$, a standard linear analysis technique can be used along with (15), (16), (24), (25) to show that $\dot{\xi}_1(t), \dot{\xi}_2(t), \dot{\xi}_3(t), \dot{\xi}_4(t) \in \mathcal{L}_\infty$ in $\mathcal{K}$. Since $\xi_1(t), \xi_2(t), \xi_3(t), \xi_4(t), \dot{\xi}_1(t), \dot{\xi}_2(t), \dot{\xi}_3(t), \dot{\xi}_4(t) \in \mathcal{L}_\infty$, (15) and (24) can be used along with the Assumption 1 to prove that $P(t)$, $V(t) \in \mathcal{L}_\infty$ in $\mathcal{K}$. Given that $P(t)$, $V(t) \in \mathcal{L}_\infty$, (9a)-(9d) along with Assumption 3 to prove that the control input $F_V(t), \tau_\Omega(t) \in \mathcal{L}_\infty$ in $\mathcal{K}$. Since $\eta_1(t), \eta_2(t) \in \mathcal{L}_\infty$, can be used to prove that $\dot{F}_V(t), \dot{\tau}_\Omega \in \mathcal{L}_\infty$ in $\mathcal{K}$. Given that $\eta_1(t), \eta_2(t), \xi_1(t), \xi_2(t), \xi_3(t), \xi_4(t) \in \mathcal{L}_\infty$, the bounding inequalities in (21), (31) can be used to prove that $\dot{\eta}_1(t), \dot{\eta}_2(t) \in \mathcal{L}_\infty$ in $\mathcal{D}$. Since $\dot{\xi}_1(t), \dot{\xi}_2(t), \dot{\xi}_3(t), \dot{\xi}_4(t), \dot{\eta}_1(t), \dot{\eta}_2(t) \in \mathcal{L}_\infty$, (23) and (33) can be used to prove that $\mu(t)$ is uniformly continuous in $\mathcal{K}$. Hence, the definitions of $U(w)$ and $\mu(t)$ can be used to prove that $U(w)$ is uniformly continuous in $\mathcal{K}$.

Let $\mathcal{S} \subset \mathcal{K}$ denote a set defined as follows:

$$\mathcal{S} \triangleq \left\{w(t) \subset \mathcal{K} | U(w(t)) \leq \frac{1}{2}\left(\rho^{-1}\left(2\sqrt{c_1 c_2}\right)\right)^2\right\}. \tag{63}$$

Theorem 8.4 of [32] can now be invoked to state that

$$c\|\mu(t)\|^2 \to 0 \quad \text{as} \quad t \to \infty \quad \forall\, w(0) \in \mathcal{S}.$$

Based on the definition of $\mu(t)$, (63) can be used to show that

$$\|\xi(t)\| \to 0 \quad \text{as} \quad t \to \infty \quad \forall\, w(0) \in \mathcal{S}. \tag{64}$$

Hence, asymptotic regulation of the quadrotor states is achieved, provided the initial conditions lie within the set $\mathcal{S}$, where $\mathcal{S}$ can be made arbitrarily large by increasing the control gain $\alpha$ - a semi-global result.

## V. SIMULATION RESULTS

In this section, extensive simulations are performed to validate the efficiency of the proposed adaptive modified RISE control algorithm for both the position and attitude loop with varying levels of uncertainty, tracking speeds, and disturbance magnitudes. We conducted a comprehensive numerical study encompassing four simulation cases with different combinations of dynamic models and control laws, as summarized in Table 1.

**TABLE 1:** Summary of simulation cases with different combinations of dynamic models and control laws. Color codes correspond to those used in the figures.

| | Dynamic Model | Control Algorithm | Color |
|---|---|---|---|
| Case-1 | Model in [26] | RISE Control [26] | Green |
| Case-2 | Model in [26] | Proposed Control | Yellow |
| Case-3 | Current Model | RISE Control [26] | Purple |
| Case-4 | Current Model | Proposed Control | Red |

**TABLE 2:** Physical Constants - Parameters

| Parameter | Description | Value |
|---|---|---|
| $K_T$ | Thrust Coefficient | $3.232\times10^{-7}\ N.s^2$ |
| $K_D$ | Drag Coefficient | $2.9842\times10^{-5}\ Nm.s^2$ |
| $[k_{d_x}, k_{d_y}, k_{d_z}]$ | Translational Drag | $[5.567, 5.567, 6.354]\times10^{-4}\,\frac{N.s}{m}$ |
| $[k_{a_x}, k_{a_y}, k_{a_z}]$ | Aerodynamic Friction | $[3.367, 6.32, 5.354]\times10^{-4}\,\frac{N.m.s}{rad}$ |

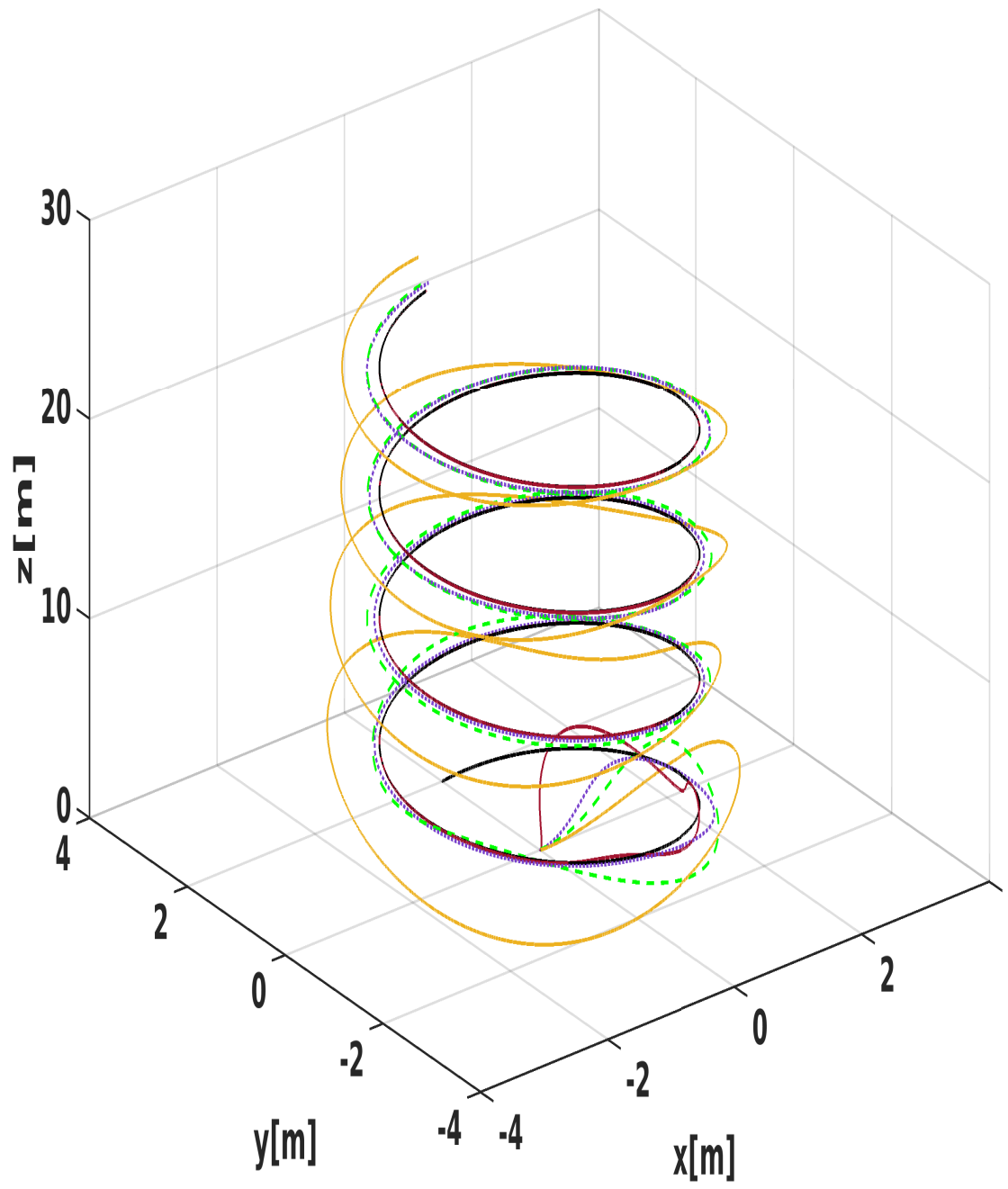

**FIGURE 1:** 3D time evolution of the position of the quadrotor for all 4 cases

In this current work, the desired trajectories are generated using the following definitions

$$\left[x_{des}, y_{des}, z_{des}, \psi_{des}\right]^T = \left[amp_d \sin(t), amp_d \cos(t), t, \sin(t)\right]^T. \tag{65}$$

The simulation results are summarized in Fig. 1-8. Fig. 1 illustrates the 3D trajectory tracking performance of all four cases. The proposed adaptive modified RISE control (Case-4, red) again demonstrates significantly improved tracking accuracy and closely follows the desired trajectory, while the other cases exhibit larger deviations, particularly under aggressive maneuvers and during fast transitions. Fig. 2 illustrates the time histories of the four individual motor speeds (for Case 3-purple and Case 4-red). As depicted, the motor speeds operate within a typical range commonly observed for quadrotor flight maneuvers [33], demonstrating that the control system's demands on the actuators remain physically realizable and consistent with standard operating conditions. Fig. 3 presents the evolution of the adaptive parameter estimates $\hat{\lambda}_1$ and $\hat{\lambda}_2$, demonstrating rapid convergence of the parameter update laws.

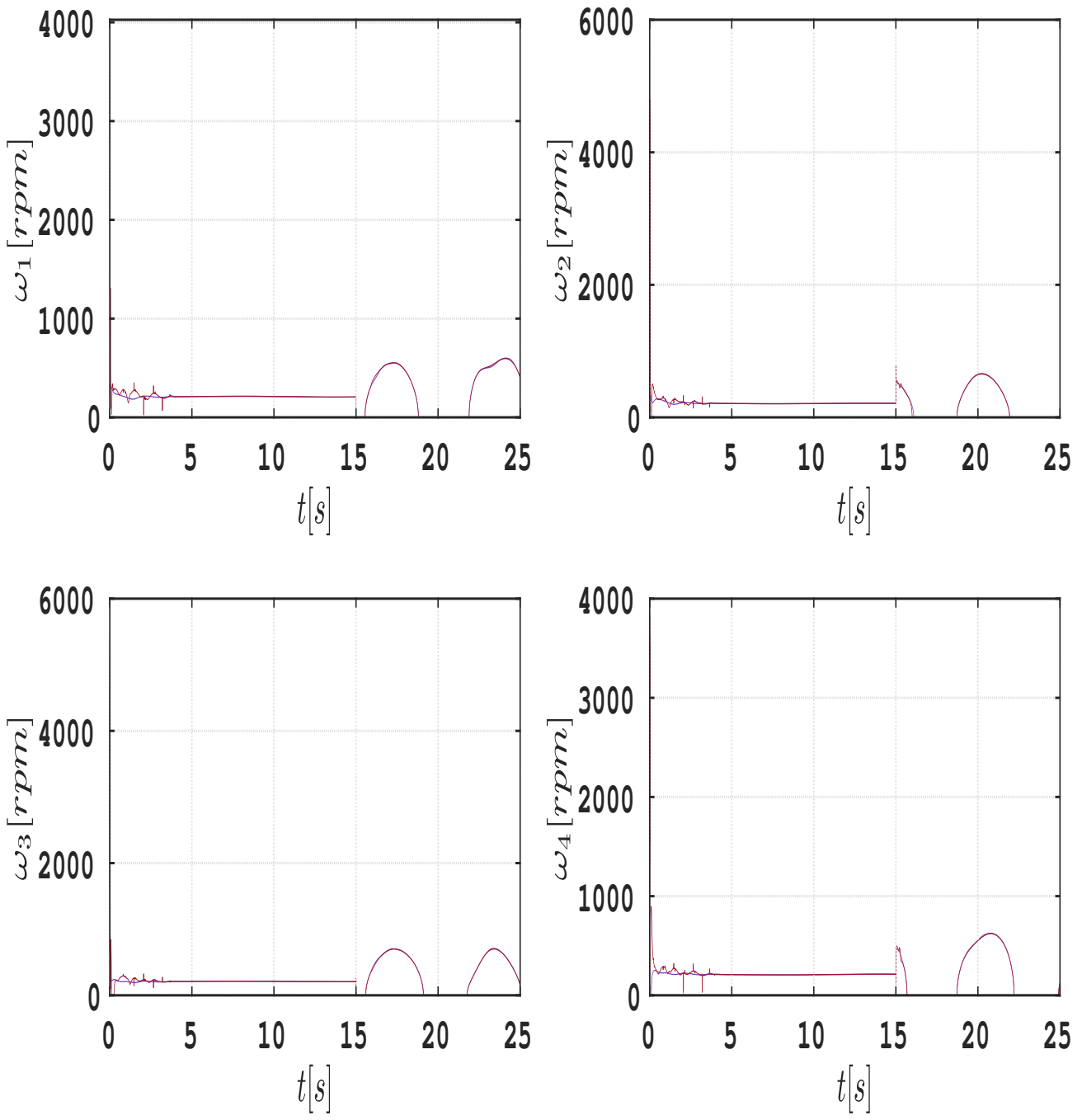

**FIGURE 2:** Time evolution of the individual motor speeds of the quadrotor during a trajectory tracking maneuver (only for case-3 and case-4)

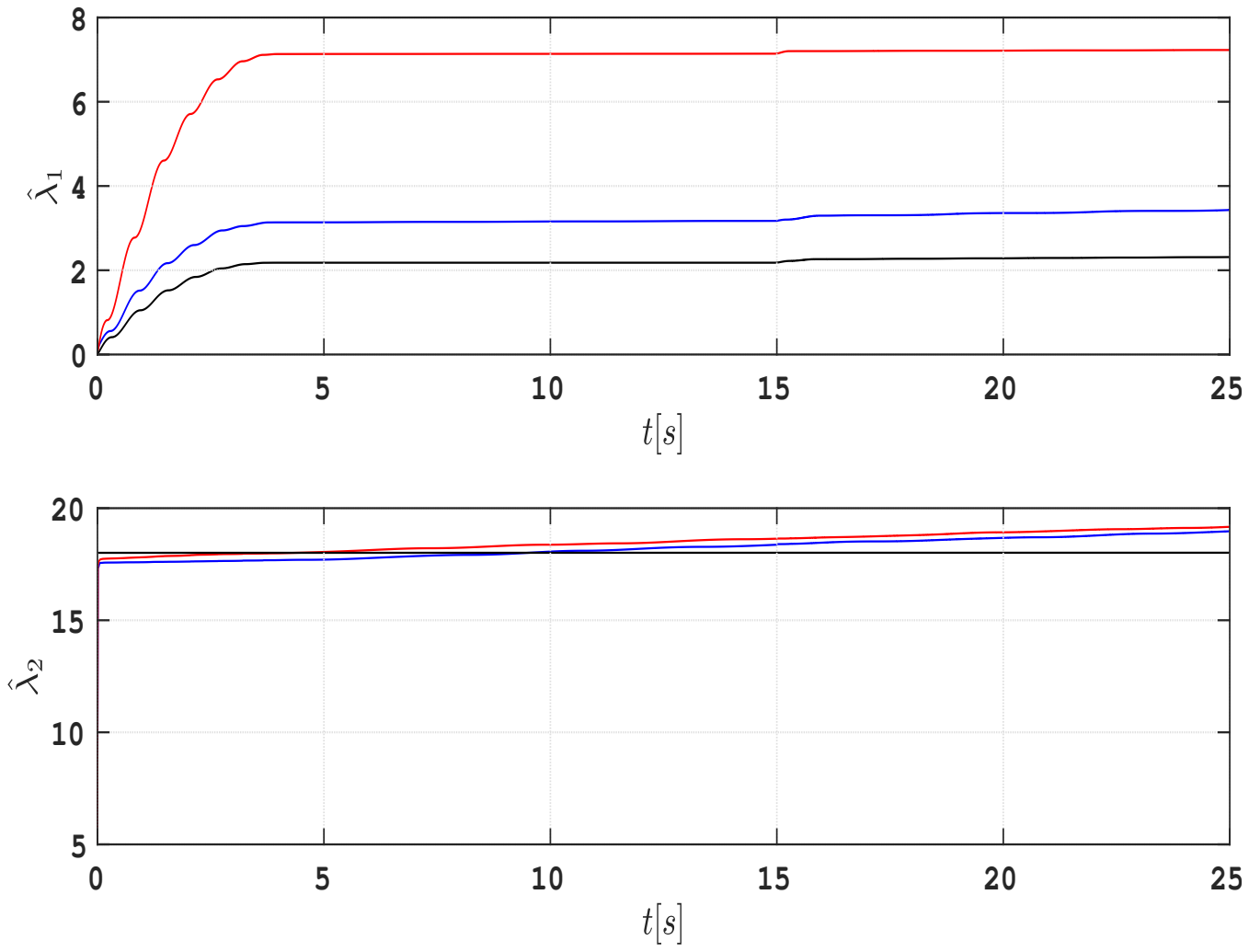

**FIGURE 3:** Time response of the adaptive parameter estimates $\hat{\lambda}_1$ and $\hat{\lambda}_2$

Fig. 4 shows the time evolution of the tracking error in transnational and rotational positions for all cases under the following conditions (uncertainty = 10%, disturbance_mag = 1, and amp_des = 2). The proposed adaptive MRISE controller (Case-4, red) consistently achieves lower tracking errors compared to baseline RISE control, particularly in the presence of model uncertainties and disturbances. Fig. 5 shows the corresponding control inputs generated during trajectory tracking. While the proposed adaptive modified RISE controller (Case-4, red) exhibits relatively high initial control magnitudes-expected due to large initial transients and model mismatch-the controller quickly settles to produce smooth and consistent control signals across all channels. This behavior is maintained under varying levels of parametric uncertainty, external disturbances, and trajectory tracking speeds. In contrast, the controllers in Cases 1-3 exhibit larger and more sustained oscillations, particularly in $F_T$ (thrust) and $\tau_\phi, \tau_\theta, \tau_\psi$ (roll, pitch, and yaw torques), which reflect their reduced robustness to system variations. Furthermore, an important practical advantage of the proposed controller is that it does not require repeated gain tuning to maintain performance across different disturbance magnitudes and tracking speeds-a challenge frequently encountered with baseline RISE control and other fixed-gain methods. This adaptability and consistent performance make the proposed control architecture highly attractive for real-world quadrotor applications.

Fig. 6-8 summarize the mean RMS error for transnational and rotational states across all four cases. Fig. 6 summarizes the root mean squared error (RMSE) in tracking transnational ($P(t)$) and rotational ($\Theta(t)$) position under varying levels of parametric uncertainty ranging from $-25\%$ to $30\%$ for all four simulation cases (See Table 1). As the uncertainty level increases, the performance gap between the controllers becomes more pronounced. Across all uncertainty levels, the proposed adaptive modified RISE controller (Case-4, red) consistently achieves the lowest RMSE in both ($P(t)$) and rotational positions, demonstrating strong robustness to parametric variations.

Fig. 7 presents the RMSE in tracking $P(t)$ and $\Theta(t)$ for all four cases under varying external disturbance magnitudes. Across all disturbance levels, the proposed adaptive modified RISE controller (Case-4, red) consistently achieves the lowest RMSE, demonstrating strong disturbance rejection capability. Fig. 8 shows the RMSE in tracking $P(t)$ and $\Theta(t)$ across all four cases as the desired trajectory speed is varied from $1\frac{m}{s}$ to $4\frac{m}{s}$ (see Eq. (65)). As the tracking speed increases, all controllers experience some degradation in tracking performance. However, the proposed adaptive modified RISE controller (Case-4, red) consistently achieves the lowest RMSE across all speeds, demonstrating superior adaptability and robustness to fast trajectory commands.

The control gains associated with the outer loop in (34) are given as $\alpha_F$ = 0.01, $\alpha_1$= 2, $k_1$ = 5, $k_2$= 3.5, and those of the inner loop in (37) are given as $\alpha_\tau$ = 0.01, $\alpha_2$= 5, $k_3$= 20, $k_4$ = 20, respectively. The initial conditions of the adaptive estimates $\hat{\lambda}_1 = [0.1, 0, 0]^T$ and $\hat{\lambda}_2 = [5, 5, 18]^T$. The initial condition of the translational state $P = [0, 0, 0]^T$, $m = 0.485$ kg, $l = 0.25m$, $g = 9.81\frac{m}{s^2}$. Based on the comprehensive results presented in Figs. 4-8, the proposed adaptive modified RISE control approach consistently outperforms the baseline methods across all evaluated scenarios. It achieves superior tracking accuracy, enhanced robustness to model uncertainties, external disturbances, and smoother control responses even at high trajectory speeds.

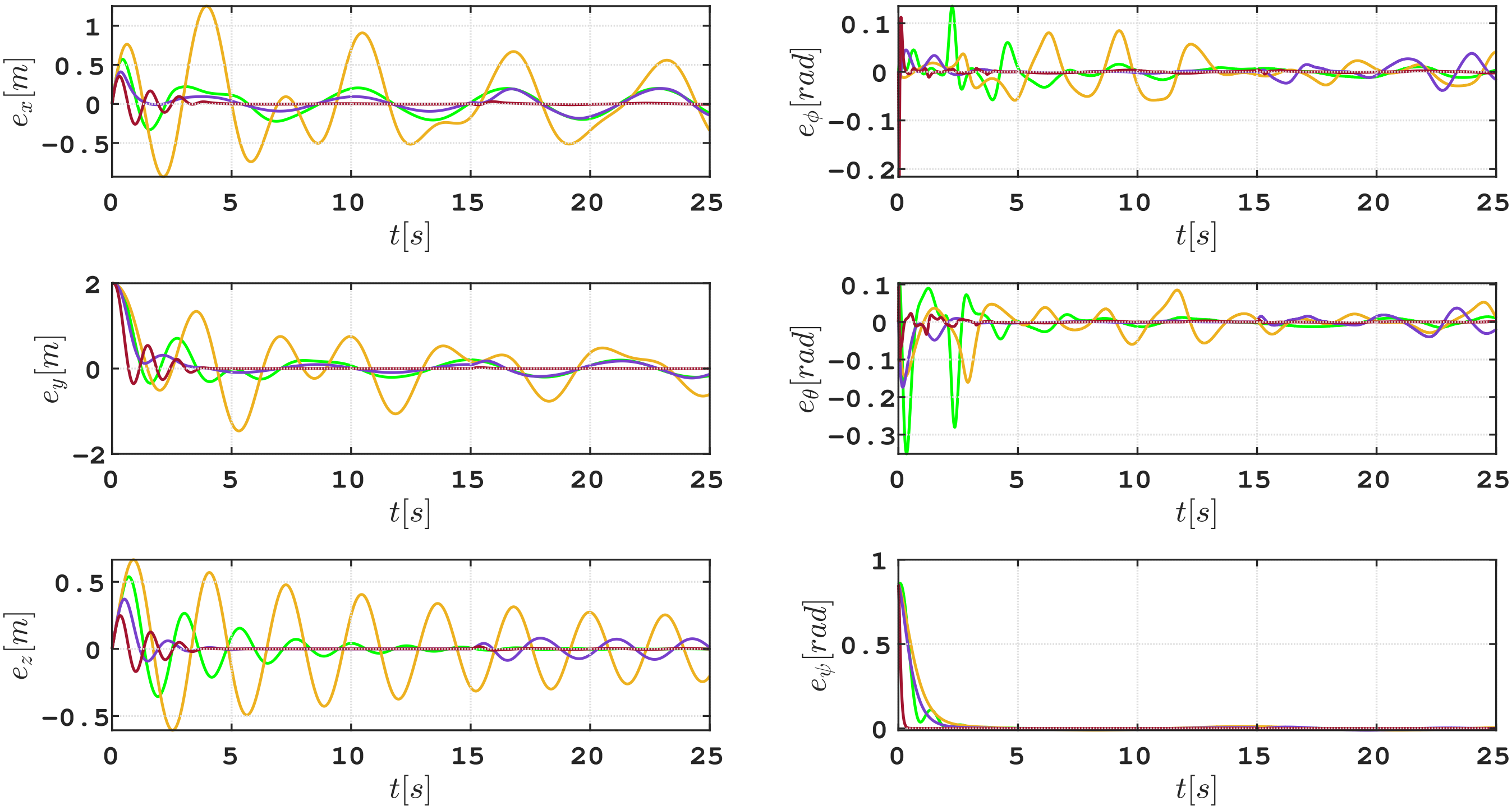

**FIGURE 4:** Time evolution of the error between the desired and measured translational position [left] and angular position [right] during the closed loop operation.

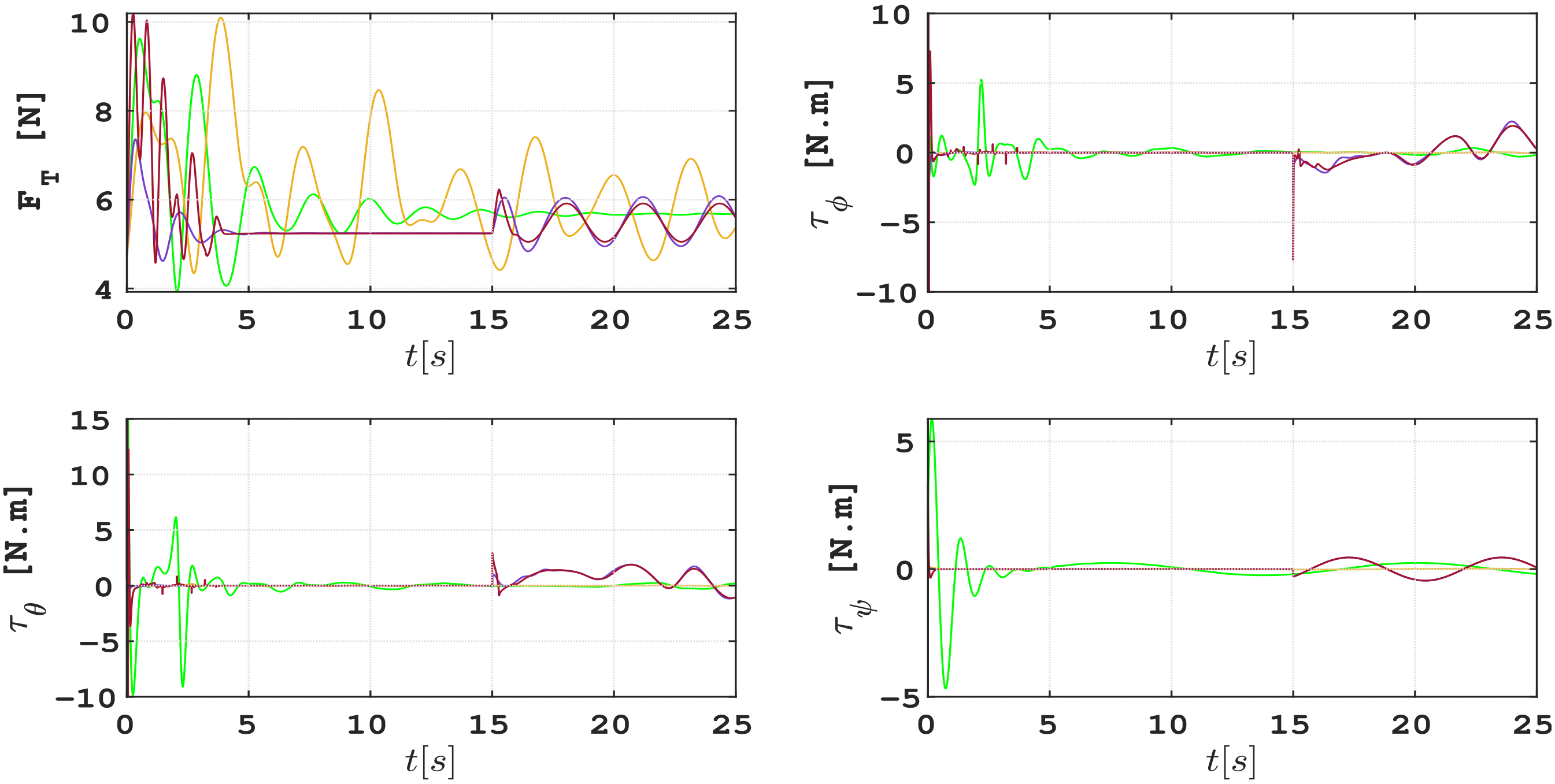

**FIGURE 5:** A zoomed in commanded control inputs during the closed-loop operation - Transient

## VI. CONCLUSION

In this paper, an adaptive modified robust inverse of signum error (AM-RISE) state feedback controller is proposed that achieves asymptotic trajectory tracking for quadrotors under gyroscopic effects, model and actuator uncertainties, and at the same time mitigating external disturbances. The control algorithm is also capable of tracking desired trajectories of various magnitudes. The proposed nonlinear control algorithms adapt to varying magnitudes of uncertainties, disturbances, and tracks trajectories of varying speeds without the need for gain re-tuning. A rigorous Lyapunov-based analysis is utilized to prove asymptotic trajectory tracking of both the position and attitude loops, where the region of convergence can be made arbitrarily large through judicious control gain selection. Detailed numerical simulations are provided to evaluate the proposed control design. Future work will focus on evaluating the proposed robust control methods through real-time experiments on a quadrotor platform (either on a CrazyFlie2.1 or ModalAI Starling).

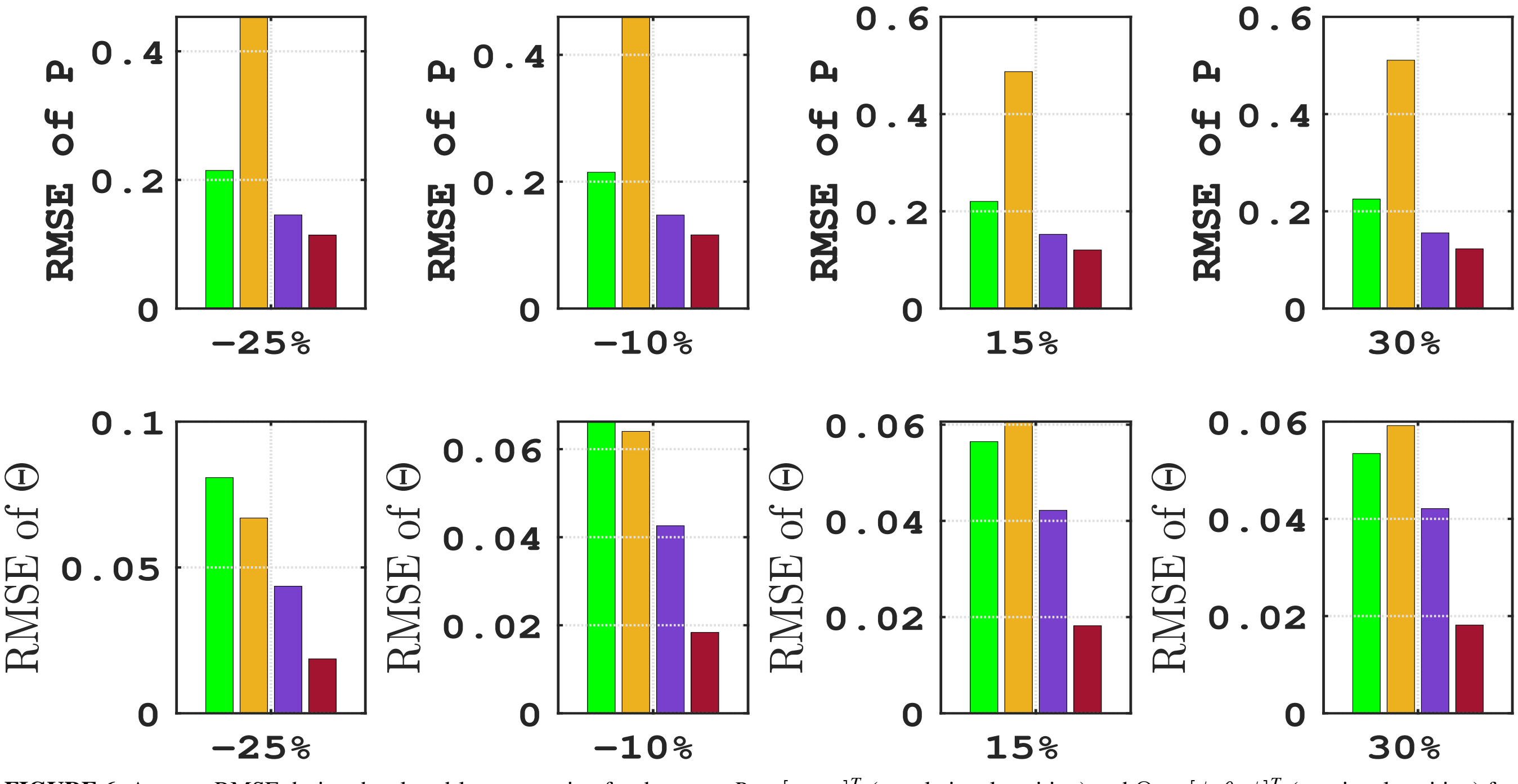

FIGURE 6: Average RMSE during the closed-loop operation for the states $P = [x, y, z]^T$ (translational position) and $\Theta = [\phi, \theta, \psi]^T$ (rotational position) for varying uncertainty levels – for all the 4 cases.

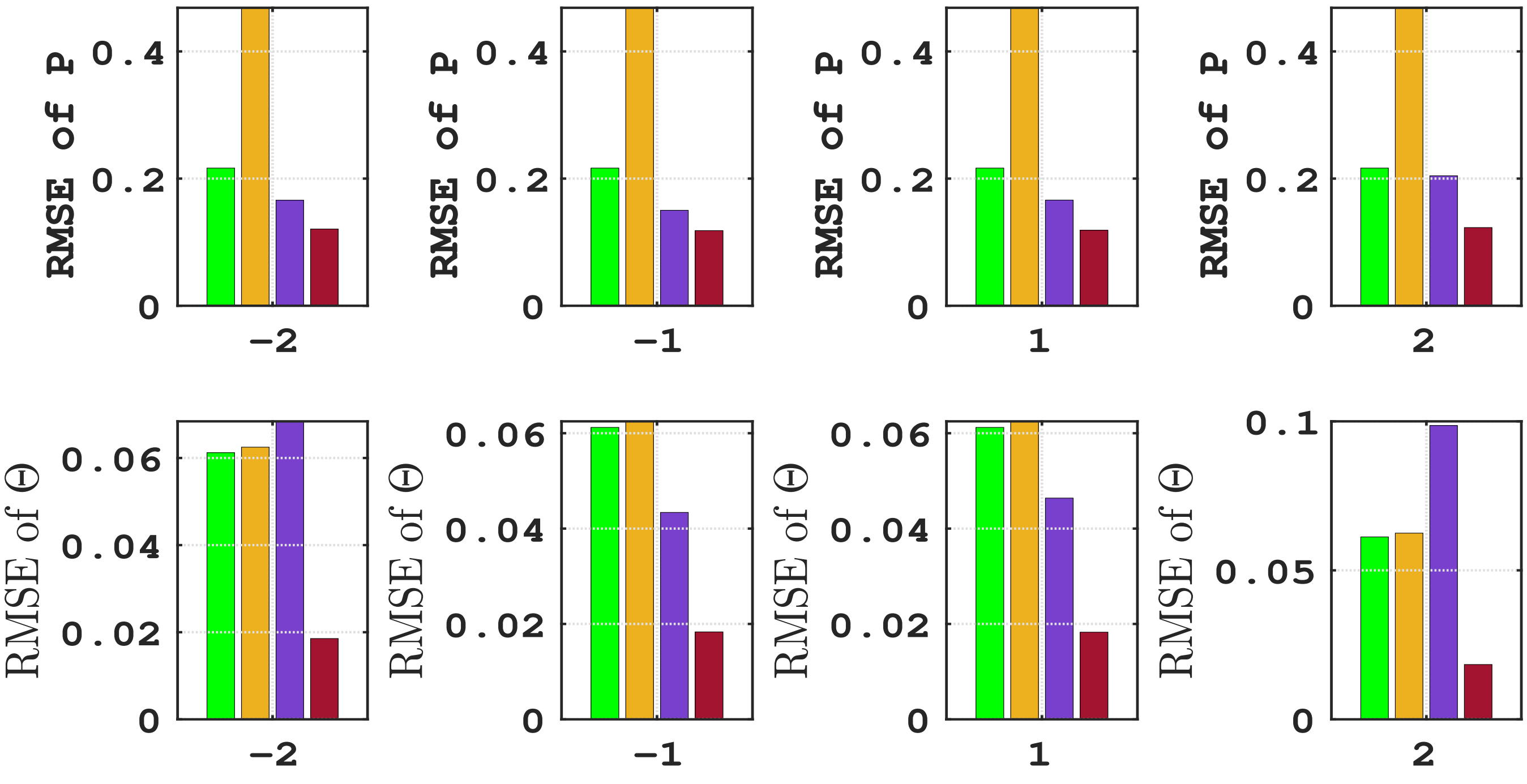

FIGURE 7: Average RMSE during the closed-loop operation for the states $P = [x, y, z]^T$ (translational position) and $\Theta = [\phi, \theta, \psi]^T$ (rotational position) for varying disturbance magnitudes - for all the 4 cases.

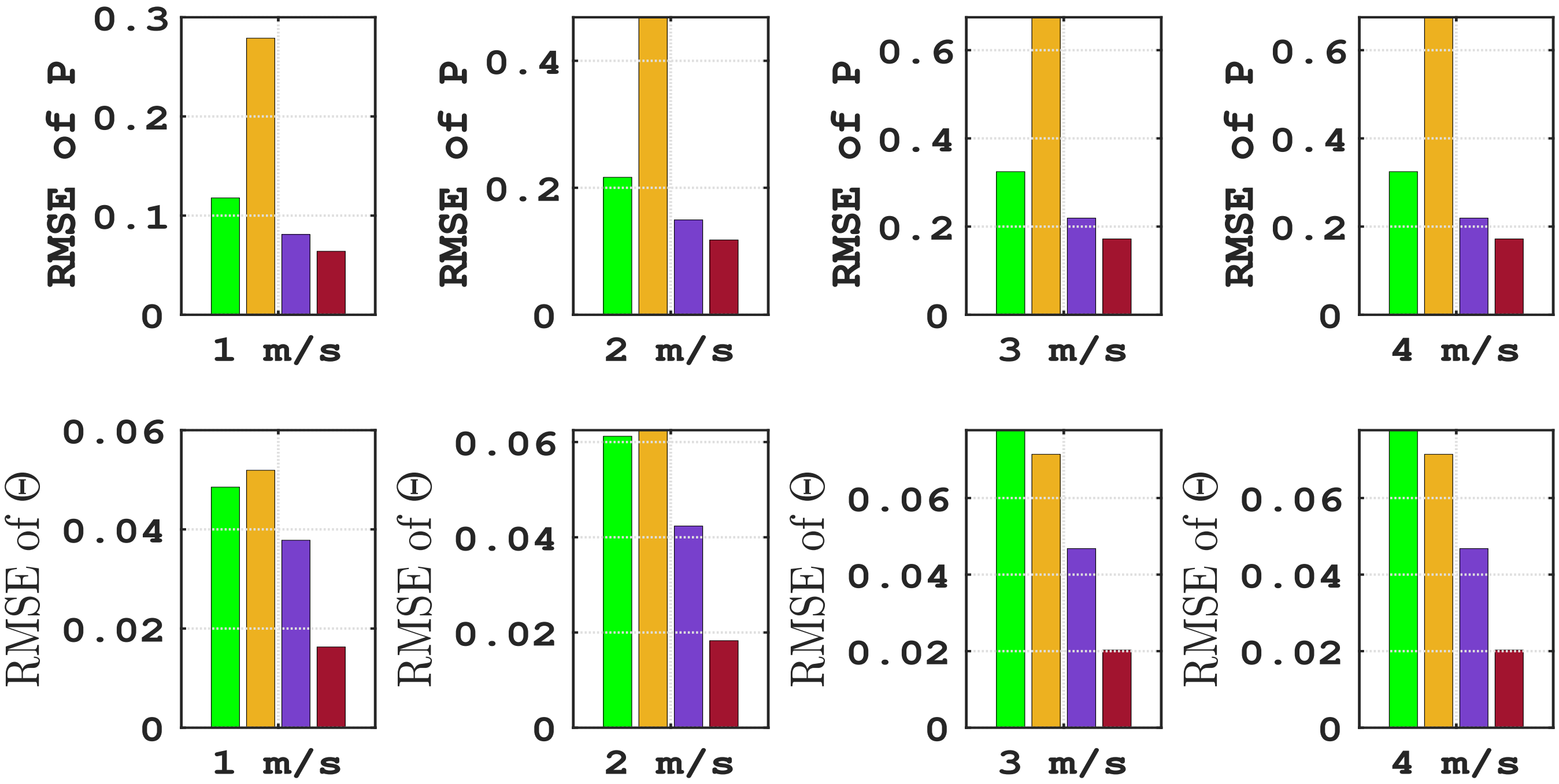

**FIGURE 8:** Average RMSE during the closed-loop operation for the states $P = [x, y, z]^T$ (translational position) and $\Theta = [\phi, \theta, \psi]^T$ (rotational position) for varying desired speeds - for all the 4 cases.

## ACKNOWLEDGMENT

The authors gratefully acknowledge the support provided by the Department of Mechanical and Aerospace Engineering, University of Dayton.